\documentclass[%
 reprint,
 amsmath,amssymb,
 aps,
pra,
]{revtex4-1}

\usepackage{graphicx}
\usepackage{dcolumn}
\usepackage{bm}

\begin{document}

\preprint{APS/123-QED}

\title{Fundamental limits for non-destructive measurement of a single spin by Faraday rotation}

\author{D. Scalbert}%
 \affiliation{Laboratoire Charles Coulomb (L2C), UMR 5221 CNRS-Universit\'{e} de Montpellier, Montpellier, FR-34095, France}

\date{\today}

\begin{abstract}
Faraday rotation being a dispersive effect, is commonly considered as the method of choice for non-destructive detection of spin states. Nevertheless Faraday rotation is inevitably accompanied by spin-flips induced by Raman scattering, which compromises non-destructive detection. Here, we derive an explicit general relation relating the Faraday rotation and the spin-flip Raman scattering cross-sections, from which precise criteria for non-destructive detection are established. It is shown that, even in ideal conditions, non-destructive measurement of a single spin can be achieved only in anisotropic media, or within an optical cavity.  


\end{abstract}


\maketitle
\textit{Introduction.}
Encoding information into the spin state of a single electron, nucleus, or atom, and reading this information non-destructively constitute some of the main challenges of quantum computing and spintronics.  These challenges have motivated strong experimental efforts towards electrical and optical detection of single spin states in semiconductors \cite{Hanson2008}. Optical detection has been demonstrated by polarized photoluminescence or polarization-dependent absorption \cite{Li2004,Bracker2005, Ebbens2005}, but these methods are destructive. Dispersive methods like non-resonant Kerr or Faraday rotation can be in principle non-destructive and open a way to quantum non-demolition measurements of a single spin state \cite{Smirnov2017}. Nevertheless it is known that even for dispersive measurements, the probe laser may eventually flip the targeted spin, compromising the non-destructive measurement. This limitation is of fundamental nature, since spin Faraday rotation is inevitably linked to spin-flip Raman scattering \cite{Shen1964, Pershan1966,Romestain1975}. This issue has been also addressed in details in the context of spin noise spectroscopy, where the signal can be interpreted either as Faraday or as Raman noise \cite{Gorbovitsky1983,Glazov2015}. To overcome this problem new schemes for quantum non-demolition measurements have been proposed \cite{Vamivakas2010, Neumann2010b,Puri2014,Delteil2014a}.  In practice the conditions for non-destructive measurements can be even more challenging to realize due to non-ideal experimental conditions, such as light scattering in the sample substrate, low detector quantum efficiency, mixing of spin states etc. Since the first detections of a single spin by Kerr or Faraday effect \cite{Berezovsky2006,Atature2007}, and thanks to strong experimental efforts, and technological progress  non-destructive measurements with these methods are within reach  \cite{Lu2009,Arnold2015}.




Until now the fundamental limits imposed by spin-flip Raman scattering (SFRS) on such measurements have not been established quantitatively. This Letter intends to fill this gap through the derivation of an explicit and general relation between the SFRS, and the spin Faraday rotation (SFR) cross-sections. The notion of cross-section is a very general and convenient way to characterize the probability of scattering, capture, or absorption, of light or particles when they interact with atoms, defects or impurities in solids. As shown in \cite{Giri2012} it is also adapted to characterize the Faraday rotation induced by a spin polarization. Instead, in general the Faraday rotation is characterized by the Verdet constant, which is the proportionality factor between the rotation angle on one side, and the interaction length times the magnetic field intensity on the other side. This definition is inconvenient for SFR because the rotation angle is proportional to the spin polarization density rather than the magnetic field. A spin polarization density can be created without applied magnetic field by optical pumping or can appear locally because of a spontaneous spin fluctuation. This property is exploited for example in spin noise spectroscopy \cite{Aleksandrov1981,Crooker2004,Oestreich2005}. In these situations it is better to introduce the proportionality factor $\sigma_\text{F}$ between the rotation angle $\theta_\text{F}$ and spin polarization density $J_z$ such that $\theta_\text{F}=\sigma_\text{F} J_z \ell$, where $\ell$ is the interaction length \cite{Giri2012}. $\sigma_\text{F}$ has the dimension of a surface and can be considered as a  cross-section for SFR.

As we show below there exists a direct and general relation between the spin-flip Raman scattering cross-section $\sigma_\text{R}$, and $\sigma_\text{F}$. This relation shows that, without an optical cavity, a quantum non-demolition of a single spin by Faraday rotation is not always feasible even in an ideal experiment.

\textit{General relations between cross-sections.}
We consider a transparent dielectric material with spins embedded in it. In these conditions the electric induction can be written as \cite{Landau1969}
\begin{equation}\label{gyrotropy}
  \mathbf{D}=\epsilon' \mathbf{E}+ i \mathbf{E}\times (G\mathbf{J}).
\end{equation}
$\epsilon'$ is the real part of the dielectric tensor, and $\mathbf{E}$ is the electric field. In general $G$ is a second rank tensor, which becomes a scalar in optically isotropic media such as atomic vapors or cubic semiconductors. $\mathbf{J}$ stands for the spin density.
We will consider below dielectric materials with point group symmetry $C_{3v}$ (relevant for crystals with wurtzite structure or for quantum dots with small in-plane asymmetry), or with cubic symmetry. With the high-symmetry axis along $z$,  $G$ takes the form
\begin{equation}\label{Gtensor}
  G=\left(\begin{array}{ccc}
            G_1 & 0 & 0 \\
            0 & G_1 & 0 \\
            0 & 0 & G_2
          \end{array} \right),
\end{equation}

In the following we will only consider a light beam propagating along $z$, and linearly polarized along $x$. The field amplitude in the dielectric is then given by $\mathbf{E}(\mathbf{r},t)=E_0 \hat{x} \exp[i(\mathbf{k}\cdot\mathbf{r}-\omega t)]$ with $\mathbf{k}=k\hat{z}$, and the light intensity in the dielectric medium is $I_0=\frac{1}{2}\epsilon_0 E_0^2 cn$ , where $n$ is the refractive index.

To calculate the Faraday rotation angle $\theta_\text{F}$ one first considers that $\mathbf{J}=J\hat{z}$ is time-independent. The solution of the wave equation is well known and gives
\begin{equation}\label{FR}
  \theta_\text{F}=\frac{\omega G_2 J \ell}{2cn\epsilon_0}
\end{equation}

From the definition of the Faraday rotation cross-section \cite{Giri2012} one obtains
\begin{equation}\label{sigmaFR}
  \sigma_\text{F}=\frac{\omega G_2}{2cn\epsilon_0}
\end{equation}

Let us now calculate the spin-flip Raman scattering (SFRS) cross-section $\sigma_R$ for non-polarized spins. For this purpose one can consider a small volume $\textit{v}$ whose dimensions are much smaller than the optical wavelength, and that contains $N$ \textit{non-interacting} spins \footnote{Since $\sigma_R$ is a property of individual spins it can be simply calculated by switching of the spin-spin interactions.}. SFRS occurs because of spin fluctuations. By definition of $\sigma_R$ the power of light scattered by the $N$ spins in $\textit{v}$ is given by $P_\text{s}=I_0 \sigma_R N$. We have thus to calculate $P_\text{s}$. In the volume $\textit{v}$ the amplitude of the spin-dependent dipole moment induced by the incident field is given by
\begin{equation}\label{dipole}
  \textbf{p}(\mathbf{r},t)= i \left(\begin{array}{c}
                           0 \\
                           -G_2E_0J_z(\mathbf{r},t)\textit{v} \\
                           G_1E_0J_y(\mathbf{r},t)\textit{v}
                         \end{array} \right)e^{i(\mathbf{k}\cdot\mathbf{r}-\omega t)}.
\end{equation}

 The total power emitted by this time-fluctuating dipole is given by
\begin{equation}\label{DipoleRadiation}
  P_\text{s}=\frac{2}{3}\frac{1}{4\pi\epsilon (c/n)^3}<|\ddot{\textbf{p}}(r,t)|^2>,
\end{equation}
where $<...>$ denotes time average.  Since the spin fluctuations are much slower than the variations of the electromagnetic field one can neglect the time-derivatives of $J_{y,z}$. Thus we get
\begin{equation}\label{PR}
  P_\text{s}=\frac{ E_0^2 n \omega^4}{12\pi\epsilon_0 c^3}\left[ G_2^2<(J_z(\mathbf{r},t)\textit{v})^2+G_1^2<(J_y(\mathbf{r},t)\textit{v})^2>\right]
\end{equation}
For $N$ independent and randomly oriented spins we have
\begin{equation}\label{SpinFluctuation}
  <(J_{y,z}(\mathbf{r},t)\textit{v})^2>=\frac{1}{3}Ns(s+1),
\end{equation}
where $s$ is the value of the individual spins.
Hence, we obtain the SFRS cross-section
\begin{equation}\label{sigmaR}
  \sigma_\text{R}=\frac{(G_1^2+G_2^2) \omega^4}{18\pi\epsilon_0^2 c^4}s(s+1).
\end{equation}
 By comparing Eqs.~(\ref{sigmaFR}) and (\ref{sigmaR}) we obtain one of the main result of this paper, which relates $\sigma_\text{R}$ and $\sigma_\text{F}$
\begin{equation}\label{SigmaRvsSigmaF}
  \sigma_\text{R}=\frac{8\pi}{9}(1+\eta)s(s+1)\left(\frac{n\sigma_F}{\lambda}\right)^2
\end{equation}
with $\eta=(G_1/G_2)^2$, and $\lambda=2\pi c/\omega$ is the light wavelength in vacuum.

It may be useful to give the expression of the differential cross section for forward (or backward) scattering. Taking into account the radiation pattern of the dipole and that only the $y$ component of the Raman dipole \footnote{See Ref. \cite{Romestain1974} for the definition of the Raman dipole.} participates to forward-scattering. Starting from Eq.~(\ref{PR}) one easily get the differential cross section

\begin{equation}\label{}
  \left(\frac{d\sigma_\text{R}}{d\Omega}\right)_0=\frac{1}{3}s(s+1)\left(\frac{n\sigma_F}{\lambda}\right)^2,
\end{equation}
in agreement with Eq.~(7) from reference \cite{Romestain1975} for $s=1/2$.

\begin{figure}
  \centering
  \includegraphics[width=8 cm]{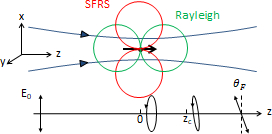}
  \caption{Up: Schematics of the geometry for dispersive measurement of a single spin by Faraday rotation. The oriented lines represent a focused gaussian beam propagating from left to right, and the arrow represents a spin at the beam waist (z=0), polarized along the $z$-axis. The green (resp. red) circles represent the radiation pattern for Rayleigh scattering (resp. SFRS). In an isotropic medium and for a spin one-half the Rayleigh and Raman scattering intensities are exactly equal. Bottom: Illustration of the effect of the Gouy phase shift on the light polarization of the gaussian beam. At $z<0$ the incident beam is linearly polarized along $x$. For $0<z\ll z_c$ the light is elliptically polarized, while for $z \gg z_c$ the light becomes linearly polarized with the polarization rotated by the angle $\theta_\text{F}$ with respect to the incoming light polarization.  }\label{FigSingleSpin}
\end{figure}

 We consider now a gaussian beam polarized along $x$ and interacting with a single spin $\mathbf{s}$ situated at the position of the beam waist $z=0$, and in a pure spin-up or spin-down state $s_z=\pm s$ (see Fig.~(\ref{FigSingleSpin})). At the beam waist $w_0$ the intensity of the field decreases with the distance $\rho$ from the beam axis as $I(\rho)=I_0\exp(-(\rho/w_0)^2)$. Since the spin is in a pure state there is no fluctuations along $z$, but only quantum spin fluctuations in the (x,y) plane. Among these, only spin fluctuations in the $y$ direction contribute to SFRS. Hence, in Eq.~(\ref{PR}) only the second term of the right-hand side must be kept, but with spin fluctuations evaluated for spin-up or spin-down state. Using $s_z=\pm s$ and $\langle s_x^2\rangle=\langle s_y^2\rangle$ one finds $\langle s_y^2\rangle=s/2$. Inserting this value in Eq.~(\ref{PR}) one obtains the following expression for the SFRS cross-section for a pure spin-up or spin-down state
 \begin{equation}\label{sigmaR_pol}
   \sigma_\text{R,pure}=\frac{4\pi}{3}\eta s\left(\frac{n\sigma_F}{\lambda}\right)^2.
 \end{equation}

Since the Raman dipole is along $z$ there is no forward SFRS in this case (see Fig.~(\ref{FigSingleSpin})), in the sense that there is no frequency-shifted scattered light in this direction. However, the static spin component along $z$ induces a dipole parallel to $y$. The emitted light is not frequency-shifted  and is cross-polarized with respect to the incident field. It corresponds to spin-Rayleigh scattering, and is at the origin of the Faraday effect. The relevant dipole associated with a pure spin state $s_z=\pm s$ is according to Eq.~(\ref{dipole}) $p_y= \mp iG_2E_0Je^{-i\omega t}$, and the emitted field along the $z$-axis can be expressed as
\begin{equation}\label{Ey}
  E_y(z,t)=\pm i\sigma_\text{F}s\frac{n}{\lambda |z|}E_0 e^{i(kz-\omega t)},
\end{equation}
where $k=n\omega/c$ is the wavevector inside the sample.
Besides, the field of the gaussian beam along the $z$ axis is given by
\begin{equation}\label{Field_gaussian}
  E_x(z,t)=E_0\frac{n\pi w_0^2}{\lambda \sqrt{1+\left(\frac{z}{z_c}\right)^2}}e^{i(kz-\omega t-\psi(r))}.
\end{equation}
where $z_c$ is the Rayleigh length, and $\psi(z)$ is the Gouy phase shift. For small rotation the Faraday rotation angle is given by $\text{Re}\left [E_y(z,t)/E_x(z,t) \right ]$. Close to the spin $\psi(z)\approx 0$, hence the Rayleigh field and the incident field are in phase quadrature. They become in phase only for $z \gg z_c$ where $\psi(z)\rightarrow \pi/2$, resulting in a rotation of the polarization plane (see Fig.~(\ref{FigSingleSpin}))
\begin{equation}\label{FR_single}
  \theta_{\text{F}\pm}=\mp\frac{\sigma_F}{\pi w_0^2}s=\mp\theta_{\text{F}}.
\end{equation}

As expected SFR is proportional to the ratio of $\sigma_\text{F}$ to the beam cross-section. Note however that the rotation angle is not constant across the beam but generally increases off-axis. Also SFR is inevitably accompanied by SFRS. Thus there is a finite probability of spin-flips induced by light, which limits the possibility of non-destructive measurement.

The total power of elastically scattered light is given by
 \begin{equation}\label{PR3}
  P_\text{el.}=\frac{G_2^2 I_0 \omega^4}{6\pi\epsilon_0^2 c^4}s^2.
\end{equation}
As for Raman scattering one gets the corresponding cross-section
 \begin{equation}\label{sigma_el}
  \sigma_\text{el.}=\frac{8\pi}{3}s^2\left(\frac{n\sigma_F}{\lambda}\right)^2,
\end{equation}
so that
\begin{equation}\label{ratioSigmaeltosigmaR}
  \frac{\sigma_\text{R,pure}}{\sigma_\text{el.}}=\frac{\eta}{2s}.
\end{equation}

\textit{Condition for non-destructive measurement of a single spin.}
\begin{figure}
  \centering
  \includegraphics[width=8 cm]{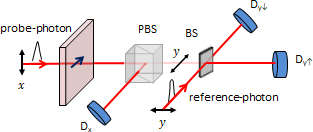}
  \caption{Principle of a single-shot measurement of a single spin via the Faraday effect using a pair of undistinguishable photons and two-photons detectors $D_{y\uparrow}$ and $D_{y\downarrow}$. $D_x$ is an optional one-photon detector.  One probe-photon is sent through the sample and interacts with the spin. The experiment is repeated until a probe-photon is transmitted through the PBS (and no photon detected on $D_x$), and interferes with a reference-photon. A first click on $D_{y\uparrow}$ or $D_{y\downarrow}$ prepares the spin in the spin-up or spin-down state. Subsequent clicks at the same detector validate the non-destructive measurement.  }\label{setup}
\end{figure}
We will consider the case of a spin one-half only. For a non-destructive measurement one demands that the acquisition time be short enough to avoid any spin-flip induced by inelastic light scattering, but long enough to determine the spin state  $|s_z\rangle=|\pm \frac{1}{2} \rangle$ from a Faraday effect based measurement. This can be done by measuring the light intensity after a linear polarizer averaged over a large number of photons (see Supplementary material). But the fundamental detection limit is more rigorously determined by looking at the level of the single spin-single photon interaction, which amounts to detect the single-photon Faraday effect \cite{Leuenberger2005, Seigneur2008}. This interaction generally leads to a spin-photon entangled state such as
\begin{equation}\label{entangled_state}
  |\Psi\rangle=\frac{1}{\sqrt{2}}\left (|-\theta_\text{F}\rangle|+\frac{1}{2}\rangle+|\theta_\text{F}\rangle|-\frac{1}{2}\rangle  \right),
\end{equation}
where $|\pm\theta_\text{F}\rangle=\cos(\theta_\text{F})|x\rangle\pm\sin(\theta_\text{F})|y\rangle$ is the photon state. Hence, a phase-sensitive detection of the $y$-component of the photon polarization state projects the spin in a pure state $|\pm\frac{1}{2}\rangle$. If the spin relaxation time is long enough, and if non-destructive measurement is achieved, the outcome of subsequent measurements should always be the same. Figure (\ref{setup}) illustrates a possible phase-sensitive detection setup, using pairs of undistinguishable photons. One photon from each pair (probe photon) is sent though the sample and interacts with the spin, while the other photon (reference photon), which is phase-stabilized with respect to the first one, is sent on the reference path. After interaction the probe photon state is projected on the $x$ and $y$ polarization states by the polarizing beam splitter (PBS). $y$-polarized probe photons then interfere with reference photons at the 50:50 beam splitter (BS).
A click, corresponding to a two-photons detection \cite{hong1987}, will occur either at detector $\text{D}_{\text{y}\uparrow}$ or at detector $\text{D}_{\text{y}\downarrow}$. This prepares the spin in a known pure state. As long as the spin state is conserved, following clicks will always occur at the same detector. At least one click is necessary to detect the spin state. The average number of clicks for $n_0$ photons incident on the sample is (assuming no loss) $n_\text{click}=n_0|\langle y|\Psi\rangle|^2\simeq n_0\theta_\text{F}^2$ for small rotation angles. Besides, the number of spin-flips due to SFRS  in the same time interval is $n_\text{sf}=n_0\sigma_\text{R,pol}/\pi w_0^2$. A non-destructive measurement requires both $n_\text{sf}\ll 1$ and $n_\text{click}\geq1$. Using Eqs.~(\ref{FR_single}) and (\ref{sigmaR_pol}) we get
\begin{equation}\label{condition_Ns}
  \frac{8\eta }{3}\left(\frac{n\pi w_0}{\lambda}\right)^2 \ll 1.
\end{equation}
Taking into account that diffraction imposes $\pi w_0\geq \lambda/n$,  we finally obtain the following condition for a non-destructive measurement
  \begin{equation}\label{condition_eta}
    \eta \ll \frac{3}{8}.
  \end{equation}
which can be satisfied only in anisotropic media. This is consistent with the fact that
for isotropic media SFRS occurs with the same probability as spin-Rayleigh scattering (see Eq.~(\ref{ratioSigmaeltosigmaR})).

We assume now that the spin is placed inside a planar microcavity. The detailed calculation of the SFRS cross-section is complicated by the modification of the optical modes in which the incident field can be scattered. We will limit ourselves to a qualitative discussion. On one hand the Faraday rotation is amplified by the quality factor of the cavity $\sigma_\text{F,cav}\simeq Q\sigma_F$ \cite{Kavokin1997,Giri2012}. On the other hand the field intensity in the cavity is, for a micro-cavity of thickness comparable to the optical wavelength inside the cavity, amplified by a comparable factor. The Raman dipole for a spin polarized along the cavity axis is also parallel to this axis, hence most of the field is scattered in directions parallel to the plane of the cavity for which the optical modes density is not modified. Therefore, in a first approximation, the SFRS cross-section is also amplified by the $Q$-factor $\sigma_\text{R,cav} \simeq Q\sigma_\text{R}$. The relation between Faraday and Raman cross-sections becomes
\begin{equation}\label{CrossSecCavity}
  \sigma_\text{R,cav}\simeq \frac{4\pi}{3} \frac{\eta s}{Q}\left(\frac{n\sigma_\text{F,cav}}{\lambda}\right)^2
\end{equation}
Hence, the condition for non-perturbative measurements becomes
  \begin{equation}\label{COnditionQ-factor}
   \frac{\eta}{Q}\ll \frac{3}{8},
  \end{equation}
which means that a non-destructive measurement is also possible in isotropic media provided the spin is placed in an optical cavity.

\textit{Application to semiconductors.}
To finish we apply the calculation to direct band gap semiconductors of cubic or wurtzite structure, with localized spin states, such as D$^{0}$X or negatively charged QDs. We note that the calculation applies also to QDs of $C_{2v}$ symmetry, in which case $G_1$ may take two different values depending on the incident light polarization with respect to the principal axis of the $G$ tensor. We assume that the optical response is dominated by the excitonic transitions towards these localized states. As is evident from Eq.~(\ref{dipole}) $G_1$ and $G_2$ are associated respectively to the $z$ and $y$-components of the Raman dipole, contributed for respectively by only light-hole transitions and by both light-hole and heavy-hole transitions. Using the well-known selection rules for these transitions we obtain
\begin{align}\label{G1etG2}
      G_1 &\propto \frac{1}{3}\frac{1}{E_\text{lh}-h\nu} \\
      G_2 &\propto \frac{1}{2}\frac{1}{E_\text{hh}-h\nu}-\frac{1}{6}\frac{1}{E_\text{lh}-h\nu},
\end{align}
where $E_{hh}$ and $E_{lh}$ are respectively the energies of the heavy-hole and light-hole excitonic transitions, and $h\nu$ is the incident photon energy. Finally,
  \begin{equation}\label{eta2}
    \eta=\frac{4}{\left[3\frac{E_\text{lh}-E_\text{hh}}{E_\text{hh}-h\nu}+2\right]^2}.
  \end{equation}
   Hence, the condition given by Eq.~(\ref{condition_eta}) will be fulfilled for large enough splitting between light-hole and heavy-hole as compared to the detuning from the heavy-hole excitonic resonance, which can be easily realized in quantum dots.

\textit{Conclusion.}
In conclusion we have derived a general relation which connects the SFRS and SFR cross-sections, valid in conditions of weak absorption. Using this relation, criteria for non-destructive measurement of a single spin state by Faraday rotation are deduced. These criteria show that non-destructive measurements require either a high enough optically anisotropy, or the use of an optical cavity. These criteria may serve as quantitative guidelines to select the experimental conditions for non-destructive measurements. The above criteria can be easily adapted for real experiments in order to take into account losses due to light scattering in the substrate and limited quantum efficiency of the detector for example.

\textit{Acknowledgements.} The author gratefully acknowledge stimulating discussions with M. Antezza, S. Cronenberger, and M. Glazov.

%


%

\end{document}